\documentclass[12pt,a4paper]{article}

\usepackage{amsfonts}
\usepackage{amsmath}
\usepackage{amsthm}
\usepackage{eucal}

\newcommand{\CC}{{\mathbb C}}
\newcommand{\DD}{{\mathbb D}}

\newcommand{\RR}{{\mathbb R}}

\newcommand{\zz}{\bar z}
\newcommand{\ZZ}{{\mathbb Z}}

\renewcommand{\epsilon}{{\varepsilon}}
\renewcommand{\phi}{{\varphi}}

\newcommand{\ch}{{\rm ch}}

\renewcommand{\Im}{{\rm Im}}
\newcommand{\Ker}{{\rm Ker}}

\newcommand{\dd}{\partial}

\renewcommand{\th}{{\rm th}}

\renewcommand{\tilde}{\widetilde}

\newcommand{\bra}{\langle}
\newcommand{\ket}{\rangle}

\newcommand{\dist}{\operatorname{dist}}

\newtheorem{thm}{Theorem}
\newtheorem{lem}[thm]{Lemma}
\newtheorem{prop}[thm]{Proposition}
\theoremstyle{remark}
\makeatletter
\renewcommand{\th@remark}{%
  \thm@headfont{\normalfont\bfseries}%
  \normalfont 
  \thm@preskip\topsep \divide\thm@preskip\tw@
  \thm@postskip\thm@preskip
}
\makeatother
\newtheorem{rem}[thm]{Remark}

\begin{document}

\title{Zero modes in a system of Aharonov--Bohm solenoids on the
  Lobachevsky plane}

\author{V.~A. Geyler\\
\emph{\normalsize Department of Mathematics, Mordovian State University}\\
\emph{\normalsize Saransk 430000, Russia}\\
\\
P. \v{S}\v{t}ov\'{\i}\v{c}ek\\
\emph{\normalsize Department of Mathematics, Faculty of Nuclear Science
}\\
\emph{\normalsize Czech Technical University}\\
\emph{\normalsize Trojanova 13, 120 00 Prague, Czech Republic}}

\date{{}}
\maketitle
\begin{abstract}
  \noindent We consider a spin $1/2$ charged particle on the
  Lobachevsky plane subjected to a magnetic field corresponding to a
  discrete system of Aharonov--Bohm solenoids. Let $H^+$ and $H^-$ be
  the two components of the Pauli operator for spin up and down,
  respectively.  We show that neither $H^+$ nor $H^-$ has a zero mode
  if the number of solenoids is finite. On the other hand, a
  construction is described of an infinite periodic system of
  solenoids for which either $H^+$ or $H^-$ has zero modes depending
  on the value of the flux carried by the solenoids.
\end{abstract}

%

\section{Introduction}
\label{sec:intro}

We consider a spin $1/2$ charged particle on the Lobachevsky plane
subjected to a time-independent magnetic field corresponding to a
discrete system of singular flux tubes perpendicular to the plane. Let
us denote by $H^+$ and $H^-$ the two components of the Pauli operator
for spin up and down, respectively. Our aim is to study zero modes in
such systems. Since both $H^+$ and $H^-$ are positive operators zero
modes are automatically ground states of the quantum system.

The current paper extends analogous results known for the Euclidean
plane to a non-flat space having a constant curvature equal to $-1$.
These results are based on the Aharonov-Casher observation \cite{AC}
that the Pauli operators for spin $1/2$ particles in a magnetic field
are related to factorisable Schr\"odinger operators. It is well known
that even in the case of a uniform magnetic field, the spectrum of the
magnetic Schr\"odinger operator $H$ changes drastically when changing
the curvature of the base plane from zero to a constant negative value
\cite{Com}. In particular, if the strength of the magnetic field is
weak enough (more precisely, if the magnetic flux through a triangle
with zero angles is less than one quantum), then the spectrum of $H$
is purely absolutely continuous in contrast to the zero curvature case
in which the spectrum is pure point. We show that the Pauli operator
with a finite number of Aharonov--Bohm fluxes exhibits a similar
behavior: it has no zero modes on the Lobachevsky plane, whereas in
the Euclidean case, the zero modes may exist in a finite system of
solenoids, as analyzed in \cite{Ara}. In this connection it is
interesting to note that the constant negative curvature exerts no
effect on the Berry phase for the zero-range potential well moving in
the uniform constant magnetic field \cite{AEG}. Furthermore, it has
been shown in \cite{GeGri} that zero modes occur if the solenoids are
arranged in an infinite plane lattice, and some generalizations and
additional details of this result can be found in \cite{GeSt},
\cite{RS}. Our Theorem \ref{theor} below is an extension of such
results to the case of the Lobachevsky plane.

As it was already mentioned, the approach we use is based on the
Aharonov--Casher ansatz. This makes it possible to employ the theory
of analytic functions when constructing the zero modes. Let us now
describe the problem in more detail and introduce the basic notation.
Some additional details related to this method are contained e.g. in
\cite{DFO}, \cite{FV}.

Let $M$ be an oriented Riemannian 2-dimensional manifold with a
conformal metric
\begin{displaymath}
  ds^2=\frac{dz\,d\zz}{\lambda^2(z,\zz)}\,,
\end{displaymath}
where $\lambda^2(z,\zz)>0$ (the function $\lambda^2(z,\zz)$ is called
the {\it Poincar\'e metric}). The corresponding area 2-form is
\begin{displaymath}
  d\sigma = \frac{dx\wedge dy}{\lambda^2(z,\zz)}
  = \frac{i}{2}\frac{dz\wedge d\zz}{\lambda^2(z,\zz)}\,.
\end{displaymath}

By definition, a magnetic field on $M$ is an exact 2-form
$b=B{}d\sigma$, where the real-valued (generalized) function $B$ is
called the {\it strength} of the field $b$. Since $b$ is exact, we
have $b=da$ where the 1-form $a=a_xdx+a_ydy=a_zdz+a_{\zz}d\zz$ is a
vector-potential of $b$. We set
$$
a_z = \frac{1}{2} (a_x-ia_y)\,,\quad a_{\zz} = \frac{1}{2}(a_x+ia_y)\,.
$$
Hence,
$$
\lambda^{-2}B = \dd_xa_y-\dd_ya_x
= \frac{2}{i}(\dd_za_{\zz}-\dd_{\zz}a_z)\,.
$$
We shall suppose that
\begin{displaymath}
  a_x\,,\,\,a_y\in L^1_{\rm loc}(M,d\sigma)\cap
  C^{\infty}(M\setminus\Omega)\,,
\end{displaymath}
for some discrete subset $\Omega$ of $M$. Moreover, we suppose that
each point of $\Omega$ is a point of discontinuity of $a_x$ or $a_y$.
Under these hypotheses, $\Omega$ is determined by $a$ in a unique way.
In particular, $a_x$ or $a_y$ may be the imaginary and the real part
of a meromorphic function, respectively.

Let us define the following operators in $L^2(M,d\sigma)$ with the
domain $C^{\infty}_0(M\setminus\Omega)$:
\begin{displaymath}
  P_x = -i\dd_x-a_x\equiv -i\nabla_x\,,\quad
  P_y = -i\dd_y-a_y\equiv -i\nabla_y\,,
\end{displaymath}
\begin{displaymath}
  \nabla_z=\frac{1}{2}(\nabla_x-i\nabla_y)=\dd_z-ia_z\,,\quad
  \nabla_{\zz}=\frac{1}{2}(\nabla_x+i\nabla_y)=\dd_{\zz}-ia_{\zz}\,,
\end{displaymath}
\begin{displaymath}
  T_{\pm}=P_x\pm iP_y=-i\nabla_x\pm \nabla_y\,.
\end{displaymath}
Let us consider the quadratic form
$$
h_{\max}^{\pm}(f) = \int\limits_M \lambda^2|T_{\pm}f|^2\,d\sigma
$$
with the domain
\begin{displaymath}
  Q(h_{\rm max}^{\pm}) = \Big\{f\in L^2(M,d\sigma);\,
  \nabla_xf,\nabla_yf\in L_{\rm loc}^1(M\setminus\Omega,d\sigma),
  \textrm{~and~~} \int\limits_M
  \lambda^2|T_{\pm}f|^2\,d\sigma<\infty\Big\}\,.
\end{displaymath}
The quadratic form $h_{\rm max}^{\pm}$ is closed and defines a
self-adjoint operator $H^{\pm}$ in $L^2(M,d\sigma)$.
On $C^{\infty}_0(M\setminus\Omega)$ we have
\begin{displaymath}
  \lambda^2T_+T_- = H^-\,,\quad \lambda^2T_-T_+ = H^+\,,
\end{displaymath}
and
\begin{displaymath}
  \lambda^{-2}H^{\pm} = P_x^2+P_y^2\mp \lambda^{-2}B\,.
\end{displaymath}
Clearly, both $H^+$ and $H^-$ are positive operators.

Suppose that in the sense of distributions
\begin{displaymath}
  \lambda^{-2}B = \frac{\dd^2\phi}{\dd x^2}+\frac{\dd^2\phi}{\dd y^2}
  \equiv \Delta\phi
\end{displaymath}
where $\phi$ is a regular distribution (a locally integrable
function). Then for the vector potential one can choose
\begin{displaymath}
  a_{\bar z} = i\partial_{\bar z}\phi,\quad
  a_{z} = -i\partial_{z}\phi,
\end{displaymath}
and the zero modes of $H^+$ (resp. $H^-$), i.e., $L^2$-solutions
$\psi\ne 0$ to the equation $H^{\pm}\psi=0$, have the form
\begin{displaymath}
  \psi(z,\zz) = \exp(\mp\phi(z,\zz))f(z,\zz)\,,
\end{displaymath}
where $f$ is a holomorphic (resp. antiholomorphic) function on
$M\setminus\Omega$.

\section{Finite number of Aharonov--Bohm solenoids}
\label{sec:finite_no}

In what follows $M$ will be the Lobachevsky plane which we shall model
as the disc
\begin{displaymath}
  \DD = \{z\in\CC;\,|z|<1\} \textrm{~with~}
  \lambda = \frac{1-z\zz}{2}\,.
\end{displaymath}
Equivalently, one could model $M$ as the upper half-plane
$\CC^+=\{z\in\CC;\,\Im\,z>0\}$ with $\lambda=(z-\zz)/(2i)$.

\begin{prop}
  Let $B$ be the magnetic field on $M$ corresponding to a finite
  family of Aharonov--Bohm solenoids with non-zero fluxes. Then
  $H^{\pm}$ has no zero modes.
\end{prop}

\begin{proof}
  Let us consider the operator $H^+$; the proof is similar in the case
  of $H^-$. Let $a_k\in\DD$, $k=1,\ldots,n$, be a finite set of
  points. Consider the function
  $$
  \phi(z,\zz) = \prod\limits_{k=1}^n|z-a_k|^{\theta_k}\,.
  $$
  Then
  $$
  \Delta \log(\phi) = 2\pi \sum\limits_{k=1}^n\theta_k\delta(z-a_k)\,,
  $$
  and the corresponding field strength equals
  $$
  B(z,\zz)=\frac{\pi}{2}
  \sum\limits_{k=1}^n\theta_k(1-|a_k|^2)^2\delta(z-a_k)\,.
  $$
  Let us note that for the field
  $\displaystyle{B=\frac{\pi}{2}\,\theta\,(1-|a|^2)^2\delta(z-a)}$ the
  flux equals
  $$
  \Phi = \frac{1}{2\pi}\int\limits_M B d\sigma=\theta\,.
  $$
  As usual, due to the gauge symmetry one can assume that
  $0<\theta_k<1$ for all $k$. Let us suppose that $H^+$ has a zero
  mode $\psi$. Then
  \begin{equation}
    \label{eq:psi_f}
    \psi(z,\zz)=\prod\limits_{k=1}^n|z-a_k|^{-\theta_k}f(z)\,,
  \end{equation}
  where $f$ is holomorphic on the domain
  $\DD\setminus\{a_1,\ldots,a_n\}$. Since $\psi\in L^2(\DD,d\sigma)$,
  the function $f$ cannot have a pole nor an essential singularity at
  any of the points $a_1,\ldots,a_n$, and therefore $f$ has an
  analytic extension to the whole domain $\DD$. Moreover, from
  (\ref{eq:psi_f}) one deduces that
  $|f(z)|\leq\textrm{const}\,|\psi(z,\zz)|$ on $\DD$ and therefore
  $f\in L^2(\DD\,,d\sigma)$. Since this means that $f^2$ is a
  holomorphic function on $\DD$ belonging to $L^1(\DD,d\sigma)$ the
  following lemma completes the proof.
\end{proof}

\begin{lem}
  \label{lem:lemma1}
  Let $f$ be a holomorphic function on $\DD$. If
  $f\in{}L^1(\DD,d\sigma)$ then $f=0$.
\end{lem}

\begin{proof}
  Let $f(z)=\sum_{n=0}^\infty{}a_nz^n$ and suppose that the series
  converges in $\DD$. Denote $z=|z|e^{i\phi}$.  The functions
  $e^{-in\phi}f(z)$ belong to $L^1(\DD,d\sigma)$ for all $n\in\ZZ$.
  Moreover, for $n\geq0$ we have
  \begin{displaymath}
    \int\limits_{\DD}e^{-in\phi}f(z)\,d\sigma
    = \lim\limits_{r\to 1-}
    \int\limits_{|z|<r}e^{-in\phi}f(z)\,d\sigma
    = 8\pi a_n\lim\limits_{r\to 1-}
    \int\limits_0^r\frac{\rho^{n+1}}{(1-\rho^2)^2}\,d\rho\,.
  \end{displaymath}
  Since the last integral diverges as $r\to1-$ it necessarily holds
  $a_n=0$.
\end{proof}

\begin{rem}
  On the Euclidean plane $\RR^2$ the following Aharonov--Casher theorem is
  valid \cite{AC}: {\it If $B(x,y)$ is a "regular" function with a compact
    support then $\dim\Ker(H^+\oplus{}H^-)=\bra|\Phi|\ket$ where
    $$
    \Phi=\frac{1}{2\pi}\int\limits_{\RR^2}B\,dxdy
    $$
    is the magnetic flux, and for $x\ge 0$,}
  $$
  \bra x\ket =
  \begin{cases}
    [x], & \textrm{if~}x\notin\ZZ, \cr
    \noalign{\medskip}
    x-1, & \textrm{if~} x\in\ZZ \textrm{~and~} x>0, \cr
    \noalign{\medskip}
    0, & \textrm{if~} x=0,
  \end{cases}
  $$
  (here $[x]$ stands for the integer part of $x$). The following
  example shows that an analogous statement is not true for the
  Lobachevsky plane.

  Let $M=\DD$ and $B(z,\zz)=\lambda^2(|z|)F(|z|)$, where
  $$
  F(r) =
  \begin{cases}
    \tilde B, & \textrm{if~} 0\le r\le r_0, \cr
    \noalign{\medskip}
    0, & \textrm{if~} r_0<r<1,
  \end{cases}
  $$
  (here $\tilde B$ is a positive number and $r_0$, $0<r_0<1$, is
  fixed). To find a function $\phi$ such that $\Delta\phi=F$ one has
  to solve the equation
  $$
  \frac{1}{r}\frac{d}{dr}r\frac{d}{dr}\phi(r) = F(r)\,.
  $$
  It is easy to show that we can set
  $$
  \displaystyle
  \phi(r) =
  \begin{cases}
    \displaystyle\frac{\tilde B}{4}r^2, & \textrm{if~}
    0\le r\le r_0, \cr
    \noalign{\medskip}
    \displaystyle\frac{\tilde B}{4}r_0^2
    +\frac{\tilde B}{2}r_0^2\log\!\left(\frac{r}{r_0}\right),
    & \textrm{if~} r_0<r<1. \cr
  \end{cases}
  $$
  It is clear that for every $\tilde{B}>0$ we have
  $$
  \inf\limits_{0\leq r\leq1}\,\exp(\mp\phi(r))>0.
  $$
  This implies that if $f\exp(\mp\phi)$ is square integrable then
  the same is true for $f$. By Lemma~\ref{lem:lemma1}, for every
  function $f\neq0$ which is holomorphic (antiholomorphic) on $\DD$ it
  holds $f\exp(\mp\phi)\notin{}L^2(\DD\,,d\sigma)$. Hence
  $\dim\Ker(H^+\oplus{}H^-)=0$. On the other hand, the flux
  $$
  \Phi = \frac{1}{2\pi}\int\limits_{\DD}Bd\sigma
  = \frac{1}{2\pi}\int\limits_{\DD}F(r)\,dxdy
  = \frac{\tilde B}{2}r_0^2
  $$
  can be an arbitrary positive number.
\end{rem}

\section{An infinite system of Aharonov--Bohm solenoids}
\label{sec:infiniteAB}

Here we consider magnetic fields with infinite total fluxes. We start
from a remark concerning a uniform magnetic field on the Lobachevsky
plane $M$.

\begin{rem}
  Suppose that $B={\rm const}$ and without loss of generality we can
  assume that $B>0$. It is known (see \cite{Com}) that in this case
  the spectrum of $H^{\pm}$ is purely absolutely continuous if and
  only if $B\le1/2$. If it is the case then the spectrum consists of
  the semi-axis $\left[1/4+B^2\mp B,\,+\infty\right[\,$. Otherwise, in
  addition to the semi-axis, the spectrum of $H^{\pm}$ contains
  infinitely degenerate eigenvalues $E_n=B(2n+1\mp1)-n^2-n$, where
  $n\in\ZZ$ and $0\le n<B-1/2$. From here one deduces that the
  operator $H^+$ has zero modes if and only if $B>1/2$ while
  $H^-\geq2B$ has never zero modes. As an illustration of the
  effectiveness of the Aharonov--Casher method let us reestablish the
  observation concerning zero modes of $H^+$.

  First, we find a function $\phi$ defined in $\DD$ such that
  $$
  \Delta \phi=B\lambda^{-2}\,.
  $$
  Assuming that $\phi$ depends on $|z|$ only we arrive at the
  equation
  $$
  \frac{1}{r}\frac{d}{dr}r\frac{d}{dr}\phi(r)=B\lambda(r)^{-2}\,.
  $$
  Its solution reads
  $$
  \phi(r) = -B\log(1-r^2)\,.
  $$
  The operator $H^+$ has a zero mode if and only if there exists a
  function $f\ne0$ which is holomorphic on $\DD$ and such that
  \begin{equation}
    \label{3.1}
    (1-r^2)^{2B}\,\frac{1}{(1-r^2)^2}\,|f(z)|^2\in
    L^1(\DD\,,dx\wedge dy)\,.
  \end{equation}
  It is clear that in the case when $B>1/2$ all functions $f$ which
  are holomorphic on $\DD$ and bounded on $\overline\DD$ satisfy
  condition (\ref{3.1}). On the other hand, suppose that a function
  $f(z)$ is holomorphic on $\DD$ and satisfies condition (\ref{3.1}).
  Denote $g(z)=f(z)^2=\sum_{m=0}^\infty{}a_mz^m$. Then for every
  $n\in\ZZ$, $n\geq0$,
  \begin{eqnarray}
    && \int\limits_{0}^{2\pi}
    \int\limits_0^1(1-r^2)^{2B}\,\frac{1}{(1-r^2)^2}\,
    g(z)e^{-in\phi}\,rdr\wedge d\phi \nonumber\\
    && \label{3.3}
    =\, 2\pi a_n\lim\limits_{\rho\to1-}\int\limits_0^\rho
    (1-r^2)^{2B}\,\frac{r^{n+1}}{(1-r^2)^2}\,dr\,.
  \end{eqnarray}
  By assumption, the integral on the LHS in (\ref{3.3}) is finite
  while the integral on the RHS converges as $\rho\to1-$ if and only
  if $B>1/2$. Hence $f(z)$ necessarily vanishes everywhere on $\DD$ if
  $B\leq1/2$.
\end{rem}

Let us recall that an action of a group $G$ on $\DD$ is called
co-compact if the factor space $\DD/G$ is compact.

\begin{lem}
  \label{lem:Lemma3}
  Let $G$ be a discrete co-compact group of isometries acting on the
  disc $\DD$ equipped with the Poincar\'e metric $ds^2$, and let $F$
  be a precompact fundamental domain for $G$. Choose an element
  $z_\gamma$ in each domain $\gamma{}F$, $\gamma\in{}G$. If $d\ge2$
  then
  \begin{equation}
    \label{3.6}
    \sum\limits_{\gamma\in G}(1-|z_\gamma|^2)^d<\infty\,.
  \end{equation}
\end{lem}

\begin{proof}
  It is sufficient to prove the lemma for $d=2$. Let us fix
  $\varepsilon$, $0<\varepsilon<1/2$. Consider a finite family
  $\{S_j\}_{j=1}^{m}$ of nonempty measurable mutually disjoint subsets
  $S_j\subset F$ such that
  \begin{itemize}

  \item[(1)] $\displaystyle \bigcup\limits_{j=1}^m S_j=F$,

  \item[(2)] ${\rm diam}\,S_j\le \epsilon,\textrm{~} \forall j$,

  \item[(3)] $\displaystyle \sigma(S_j)=\frac{1}{m}\,\sigma(F)$,

  \end{itemize}
  ($\sigma$ stands for the area). Denote by $m_{j\gamma}$ (resp.
  $M_{j\gamma}$) the infimum (resp. the supremum) of the function
  $h(z,\zz)=(1-|z|^2)^2$ on the set $\overline{\gamma S_j}$. It is
  sufficient to verify that
  $$
  \sum\limits_{j=1}^m\sum\limits_{\gamma\in G}M_{j\gamma}<\infty\,.
  $$

  It is convenient to employ the polar geodesic coordinates
  $(\rho,\theta)$ on $\DD$ centered at $z=0$. If $z=re^{i\phi}$ then
  $$
  \displaystyle
  r = \th\!\left(\frac{\rho}{2}\right),\textrm{~}\phi = \theta\,.
  $$
  In these coordinates,
  $$
  h(\rho,\theta) = \ch\!\left(\frac{\rho}{2}\right)^{\!-4}.
  $$
  From the triangle inequality it follows that for any couple of
  points from $\DD$ it holds
  \begin{displaymath}
    |\rho_1-\rho_2|
    \leq \dist\!\big((\rho_1,\theta_1),(\rho_2,\theta_2)\big)
  \end{displaymath}
  (where $\dist(\cdot,\cdot)$ is the distance in the Lobachevsky
  plane) and therefore
  \begin{displaymath}
    \sup\{|\rho_1-\rho_2|;\textrm{~}
    (\rho_1,\theta_1),(\rho_2,\theta_2)\in \gamma S_j\}
    \leq \varepsilon.
  \end{displaymath}
  Since $h$ is independent of $\theta$ we have
  \begin{eqnarray*}
    M_{j\gamma}-m_{j\gamma}
    &\leq& \epsilon\,\sup\left\{\left|\frac{d}{d\rho}\,
        \ch\left(\frac{\rho}{2}\right)^{\!-4}\right|;\textrm{~}
      (\rho,\theta)\in\gamma S_j\right\} \\
    &=& 2\epsilon\,\sup\left\{
      \ch\!\left(\frac{\rho}{2}\right)^{\!-4}
      \th\!\left(\frac{\rho}{2}\right);\textrm{~}
      (\rho,\theta)\in\gamma S_j\right\} \\
    &\leq& 2\epsilon\,\sup\left\{
      \ch\!\left(\frac{\rho}{2}\right)^{\!-4};\textrm{~}
      (\rho,\theta)\in\gamma S_j\right\} \\
    &=& 2\epsilon M_{j\gamma}\,.
  \end{eqnarray*}
  Consequently,
  $$
  M_{j\gamma}\le \frac{m_{j\gamma}}{1-2\epsilon}\le
  \frac{m}{(1-2\epsilon)\sigma(F)}\,
  \int\limits_{\gamma S_j}\,h(\rho,\theta)\,d\sigma
  $$
  and so
  \begin{equation}
    \label{3.8}
    \sum\limits_{j\gamma}M_{j\gamma}\le
    \frac{m}{(1-2\epsilon)\sigma(F)}\,
    \int\limits_{\DD}\,h(\rho,\theta)\,d\sigma
    = \frac{4m\pi}{(1-2\epsilon)\sigma(F)}\,.
  \end{equation}
  This proves the lemma.
\end{proof}

\begin{rem}
  If the points $z_\gamma$ are congruent modulo $G$ then inequality
  (\ref{3.6}) is well known and it is true for every discrete group
  $G$ (see \cite[Lemma~III.5.2]{Kra}).
\end{rem}

\begin{rem}
  Let $K=-1$ be the Gaussian curvature of the Lobachevsky plane and
  let $g$ be the genus of the closed surface $\DD/G$. The
  Gauss--Bonnet formula tells us that
  \begin{displaymath}
    \frac{1}{2\pi}\int_{\DD/G}K\,d\sigma
    = -\frac{1}{2\pi}\,\sigma(F) = 2-2g.
  \end{displaymath}
  Hence $g\geq2$ and we have $\sigma(F)\geq4\pi$ independently of the
  group $G$. Moreover, we can choose
  $$
  m = \left[\frac{\sigma(F)}{\epsilon}\right]+1\,.
  $$
  With this choice the RHS of (\ref{3.8}) can be further estimated
  from above by the expression
  \begin{displaymath}
    \frac{1}{1-2\varepsilon}\left(\frac{4\pi}{\varepsilon}
    +1\right)
  \end{displaymath}
  which is already independent of $G$. In particular, for
  $\varepsilon=1/4$ we get the upper bound $32\pi+2$. In the case of
  arbitrary $d\ge 2$ we have the estimate
  $$
  \sum\limits_{\gamma\in G}(1-|z_\gamma|^2)^d
  <\frac{4m\pi}{(1-d\epsilon)(d-1)\sigma(F)}\,,
  $$
  where $\epsilon<1/d$ and the RHS can be again replaced by an
  expression independent of $G$.
\end{rem}

Recall that the group of motions of $\DD$ regarded as the Lobachevsky
plane is $SU(1,1)$, the group of transformations
$$
Az = \frac{az+b}{\bar bz+\bar a}\,, \quad{\rm where}\quad
|a|^2-|b|^2=1\,.
$$
Let $G$ be a discrete co-compact subgroup of $SU(1,1)$ and let $F$
be a precompact fundamental domain of $G$. Suppose that $W(z)$ is an
automorphic form on $\DD$ of weight $2k$, $k\ge1$, with respect to
$G$, i.e., $W(z)$ is a meromorphic function on $\DD$ obeying the
following condition:
\begin{equation}
  \label{eq:W_automorphic}
  \forall A\in G,\textrm{~}W(Az) = A'(z)^{-k}\,W(z).
\end{equation}
For simplicity we restrict ourselves to the case when $W$ has only
simple poles and zeroes. Let us note that if $G$ is a discrete group
then automorphic forms do indeed exist, see for example
\cite[Chp.~III]{Kra}.

We can choose $F$ in such a way that $\dd{}F$ contains no poles nor
zeroes of $W$. Let $a_1,\ldots,a_n$ be the set of all zeroes and let
$b_1,\ldots,b_m$ be the set of all poles of $W$ in $F$. It is known
that $n>m$ (see \cite[\S49,~Theorem 4]{For}). Then the function
$B=\theta\lambda^{-2}\Delta\log(|W|)$, $\theta\in\RR$, is the strength
of the magnetic field of a system of Aharonov-Bohm solenoids
intersecting the Lobachevsky plane at the points $\gamma a_j$ and
$\gamma b_j$ where $\gamma$ is an arbitrary transformation from $G$.
A solenoid intersecting the plane at $\gamma a_j$ carries the flux
$\theta$, and a solenoid intersecting the plane at $\gamma b_j$
carries the flux $-\theta$.

Using the gauge symmetry we again assume, without loss of generality,
that $0<\theta<1$.

\begin{thm}\label{theor}
  If $k\theta\ge1$ then the operator $H^+(B)$ has zero modes. If
  $0<k\theta<k-1$ then the operator $H^-(B)$ has zero modes.
\end{thm}

\begin{proof}
  We restrict ourselves to the case of operator $H^+$; the proof is
  similar for $H^-$. To prove the claim one has to find a function
  $f(z)$ analytic in $\DD$ such that the function
  $$
  \psi(z,\zz)=f(z)\,|W(z)|^{-\theta}
  $$
  belongs to $L^2(\DD\,,d\sigma)$.

  One can easily check that
  \begin{equation}
    \label{eq:lambda_automorph}
    \forall A\in SU(1,1),\textrm{~}
    \lambda(Az,\overline{Az}) = |A'(z)|\,\lambda(z,\zz).
  \end{equation}
  From (\ref{eq:W_automorphic}) and (\ref{eq:lambda_automorph}) it
  follows that
  $$
  |W(z)|=(1-|z|^2)^{-k}\,r(z,\zz)
  $$
  where $r(z,\zz)$ is a $G$-periodic function. Hence
  \begin{displaymath}
    |W(z)|^{-2\theta}
    = (1-|z|^2)^{2k\theta}\,\,r(z,\zz)^{-2\theta}\,,
  \end{displaymath}
  It is clear that $r^{-2\theta}\in L^1(F,d\sigma)$ ($W(z)$ has
  only simple zeroes and so the singularities of
  $r(z,\zz)^{-2\theta}$ are integrable). Consequently, for every
  function $f$ which is bounded and analytic on $\DD$ we have
  \begin{eqnarray*}
    \int\limits_{\DD}|f(z)|^2\,|W(z)|^{-2\theta}\,d\sigma
    &=& \sum\limits_{\gamma\in G}\,\,\,
    \int\limits_{\gamma F}|f(z)|^2\,(1-|z|^2)^{2k\theta}\,
    r(z,\zz)^{-2\theta}\,d\sigma \\
    &\leq& \|f\|_{\infty}
    \int\limits_F\,r(z,\zz)^{-2\theta}\,d\sigma\,
    \sum\limits_{\gamma\in G}(1-|z_\gamma|^2)^{2k\theta}
  \end{eqnarray*}
  where $z_\gamma$ is a point from $\overline{\gamma F}$. By
  Lemma~\ref{lem:Lemma3},
  $\sum_{\gamma\in{}G}(1-|z_\gamma|^2)^{2k\theta}<\infty$. This
  completes the proof.
\end{proof}

\section*{Acknowledgments} V.~G. was supported by Grants of
DFG--RAS and INTAS.  P.~\v{S}. wishes to acknowledge gratefully the
support from the grant No. 201/05/0857 of Grant Agency of the Czech
Republic. The first named author is also grateful to the Czech
Technical University for the warm hospitality during the preparation
of this article.


\end{document}